\documentclass[aps, pra, showpacs, twocolumn, amsfonts, amsmath, mathptmx,mathrsfs,frcursive,amssymb, superscriptaddress, float]{revtex4} 

\usepackage{mathrsfs}
\usepackage{amsmath}
\usepackage{graphicx}

\begin{document}

\title{Probing quantum transport by engineering correlations in a
  speckle potential}

\author{Ardavan Alamir}

\affiliation{Universit\'e de Nice - Sophia Antipolis, Institut non Lin\'eaire de Nice, CNRS, 1361 route des Luciole
s, 06560 Valbonne, France}

\author{Pablo Capuzzi} 

\affiliation{Departamento de F\'{\i}sica, Facultad de Ciencias Exactas
  y Naturales, Universidad de Buenos Aires, 1428, Buenos Aires,
  Argentina} 

\affiliation{Instituto de F\'{\i}sica de Buenos Aires --
  CONICET, Argentina} 

\author{Samir Vartabi Kashanian}
\affiliation{Universit\'e de Nice - Sophia Antipolis, Institut non
  Lin\'eaire de Nice, CNRS, 1361 route des Luciole s, 06560 Valbonne,
  France}
\affiliation{Observatoire de la C\^ote d'Azur (ARTEMIS), Universit\'e de Nice-Sophia Antipolis, CNRS, 06304 Nice, France}

\author{Patrizia Vignolo}

\affiliation{Universit\'e de Nice - Sophia Antipolis, Institut non
  Lin\'eaire de Nice, CNRS, 1361 route des Luciole s, 06560 Valbonne,
  France}

\begin{abstract}
  We develop a procedure to modify the correlations of a speckle
  potential.  This procedure, that is suitable for spatial light
  modulator devices, allows one to increase the localization efficiency of
  the speckle in a narrow energy region whose position can be easily
  tuned.  This peculiar energy-dependent localization behavior is
  explored by pulling the potential through a cigar-shaped
  Bose-Einstein condensate.  We show that the percentage of dragged
  atoms as a function of the pulling velocity depends on the potential
  correlations below a threshold of the disorder strength. Above this
  threshold, interference effects are no longer clearly observable
  during the condensate drag.

\end{abstract}

%03.75.Kk=Dynamic properties of condensates; collective and hydrodynamic excitations, superfluid flow 
%67.85.De=(Ultracold gases, trapped gases)Dynamic properties of condensates; excitations, and superfluid flow  
%71.23.An=(Electronic structure of disordered solids.) Theories and models; localized states 
 
\maketitle
\section{Introduction}
The interplay between disorder and interactions in many-body systems
gives rise to a remarkable richness of phenomena.  In the absence of
interactions, the presence of a random potential induces the
suppression of wave propagation, as predicted by Anderson
\cite{Ande58,Ande85}. In Anderson localization, the waves diffracted
by the impurities interfere destructively in the forward direction,
with a resulting vanishing wave transmission and exponentially
localized eigenstates.  On the other hand, in the absence of disorder,
interactions can induce localized states such as gap solitons
\cite{Eiermann2004}, and suppress transport as in the Mott regime
\cite{Fisher1989}.

If an interacting quantum gas is subjected to a disorder potential,
exotic phases appear on lattice
systems~\cite{Fisher1989,Scalettar1991,Ristivojevic2012}.  In
continuum systems, it was shown that disorder shifts the onset of
superfluidity to lower \cite{Pilati2009,Allard2012} or larger
\cite{Pilati2009} critical temperatures.  In the superfluid regime,
the presence of a random potential does not perturb the dynamics of
the system in the low-energy regime.  Indeed, below a critical
velocity $v_{cr}$ that depends on the gas density and on the disorder
strength \cite{Onofrio2000,Astrakharchik2004,Ianeselli2006}, the
system, being superfluid, does not scatter against the potential
defects.  On the contrary, at velocities greater than $v_{cr}$,
superfluidity breaks down and the interference of the scattered waves
may deeply modify the system transport \cite{leboeuf07,leboeuf09} unto
the Anderson localization regime.

The authors of Refs. \cite{leboeuf07,leboeuf09} studied the transport
of a homogeneous one-dimensional (1D) interacting Bose-Einstein
condensate (BEC) in the presence of a moving random potential of
finite extent $L$.  They proved the presence of an Anderson
localization regime by studying the transmission of the BEC through
the potential and showing that it decays exponentially with $L$.
However, in ordinary ultracold-atom experiments, BECs are trapped in a
harmonic confinements and thus they are inhomogeneous.  Transmission
is no longer a well defined observable in such a geometry, however one
can identify the presence of some localization effects by studying the
time evolution of the BEC center-of-mass
\cite{Alamir2012,Alamir2013}. If the center-of-mass follows the moving
random potential, the BEC is trapped by the random potential; it
remains difficult to say if this localization is classical or induced
by the interference of the scattered fluid.

In this paper we show that it is possible to enhance the role of
interference in the localization process of an inhomogeneous
interacting BEC by introducing tunable correlations in the disorder
potential.  Our reference potential is the speckle since it is the
paradigm of the disordered potentials in ultracold atom experiments
\cite{Clem05,Billy2008,Kondov2011,Jendrzejewski2012,Shapiro2012}. The
spectral function of a conventional speckle has a finite-$k$ support
and decreases monotonically with the energy. In this work, we propose
a novel speckle whose spectral function is also defined on a compact
space but which posseses a narrow peak whose energy position is easily
tuned by varying just one setup parameter.  Our scheme, that is
illustrated in Sec. \ref{speckle}, can be straight implemented with a
Spatial Light Modulator (SLM) device.

As shown in Sec. \ref{loc}, a peak in the spectral function results
in a peak of the single-particle localization efficiency at a given
energy, meaning that high-energy particles can be localized in a
selective way.  This is crucial in our setup where one needs to exceed
the threshold $v_{cr}$ of the pulling velocity of the random potential
to break down superfluidity and observe Anderson localization
\cite{leboeuf07,leboeuf09}. Thanks to the versatility of our
potential, it is possible to drive the efficiency of the localization
toward this energy range, and then to study the BEC localization as a
function of the energy by varying the relative velocity between the
BEC and the random potential. The observation of a localization peak
in the expected energy range is a clear signature of the role of
interference, and thus of the quantum nature, in the localization
process of the boson gas.

The paper is organized as follows. In Sec. \ref{speckle} the
experimental proposal for the realization of our unconventional
speckle is illustrated and its statistical properties are
analyzed. The single-particle localization efficiency of a potential
realized with this speckle is studied in Sec. \ref{loc}.  In Sec.
\ref{model} we introduce the time-dependent nonpolynomial nonlinear
Schr\"odinger equation (NPSE) that describes the condensate dynamics
in the elongated geometry and in the presence of a moving disorder
potential.  In Sec. \ref{results} we show that the localization
efficiency of the random potential depends on the correlations of the
potential only at small values of the disorder strength. At larger
potential strength, the percentage of localized atoms is no longer
sensitive to the microscopic details of the disorder: the BEC is just
classically trapped by the potential wells.  Our concluding remarks
are given in Sec. \ref{concl}.

\section{Speckle potential with tunable correlations}
\label{speckle}
\begin{figure*}
\includegraphics[width=0.9\linewidth]{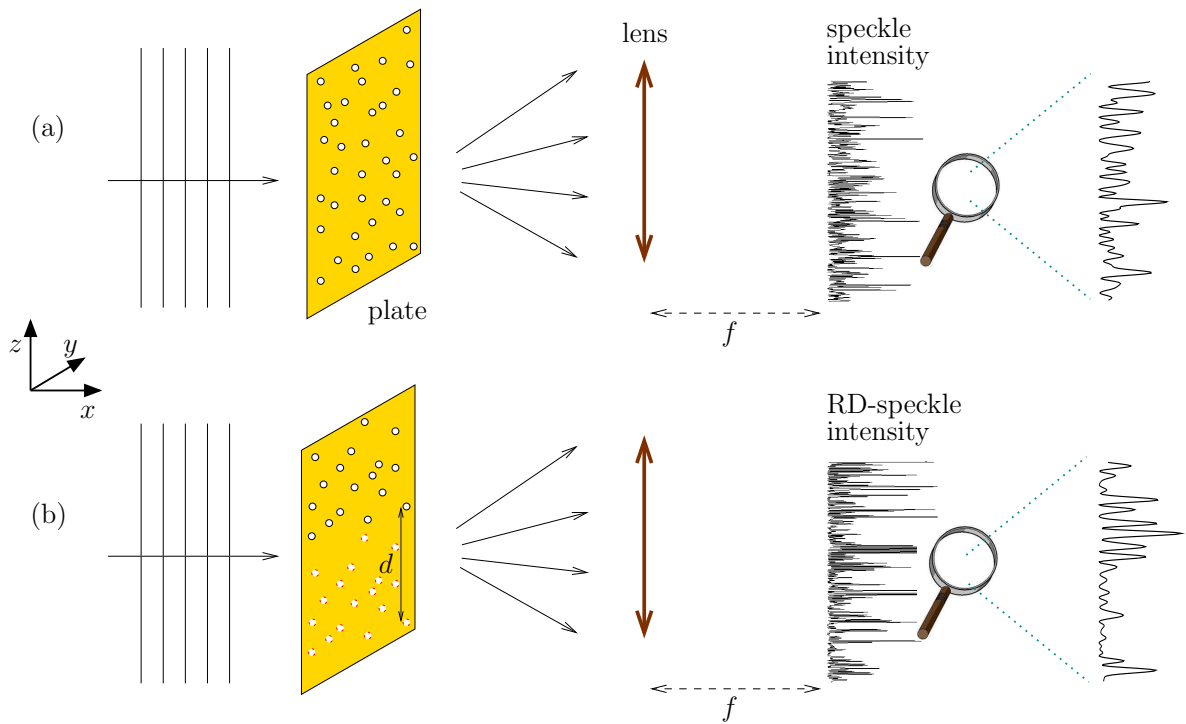}
\caption{\label{fig-schema} (Color online) Schematic representation of
  the device allowing tunable correlations in the speckle: an incident
  plane wave is diffracted by a plate with a random distribution of
  holes. The diffraction pattern obtained in the focal plane of a
  converging lens is a standard speckle if the hole distribution is
  $\delta$-correlated (a), while it is a RD-speckle if the hole
  distribution is dimerized in the $z$ direction (b). In the plate in
  the bottom row, the different borders of the hole (black continuous
  line and red dotted line) are a guide for the eyes to identify the
  dimerization of the hole distribution.}
\end{figure*}
\begin{figure}
\includegraphics[width=0.75\linewidth]{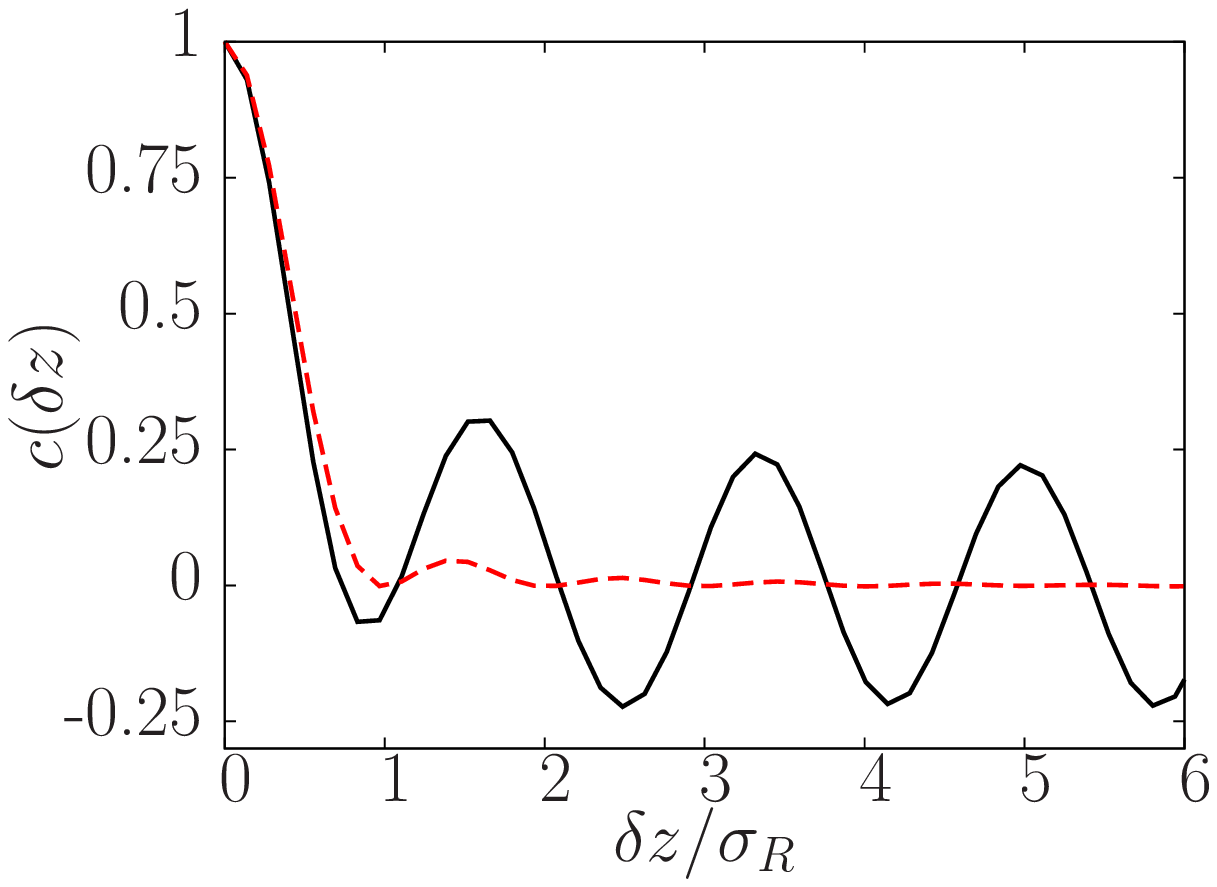}
\includegraphics[width=0.75\linewidth]{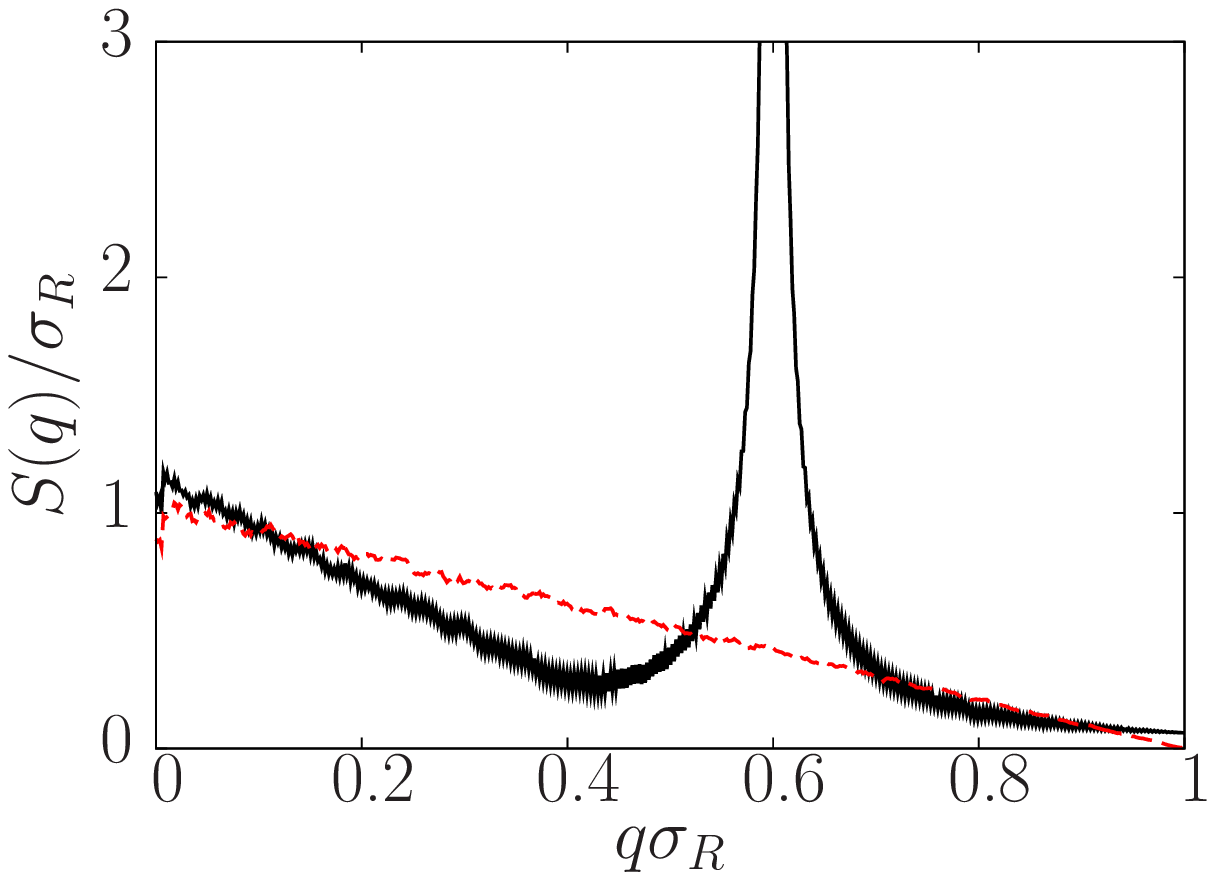}
\caption{(Color online)\label{fig-corr} Rescaled correlation function
  $c(\delta z)$ as a function of $\delta z$ in unit of $\sigma_R$ (top
  panel) and spectral function $S(q)$ (in units of $\sigma_R$) as a
  function of $q$ in unit of $1/\sigma_R$ (bottom panel) for the case
  of a standard speckle (dashed-red line) and a RD-speckle
  (continuous-black line).  We set $L=2$ cm, $\lambda_L = 532$ nm,
  $f=2.3$ cm, $d=1.20$ cm, and $N_{h}$ is of the order of 100. The
  results were calculated from the random potentials of finite extent
  used in the dynamical calculations of Sec. \ \ref{model}, averaging
  over 500 configurations.}
\end{figure}

To generate a speckle, we consider the set-up illustrated in
Fig. \ref{fig-schema}(a).  An incident plane wave of wavelength
$\lambda_L$ is diffracted by a matte square plate of side $L$ covered
with a random distribution of $N_h$ identical holes of radius $r$.
The Fraunhofer diffraction pattern obtained in the focal plane of a
converging lens of focal length $f$ is given by
\begin{equation}
I(y,z)=I_{h}(y,z)\left|\sum_{i=1}^{N_h} e^{-\frac{2i\pi}{\lambda_L f}(yy_i+zz_i)}
\right|^2
\end{equation} 
where $I_{h}(y,z)$ is the diffraction pattern of a single hole, and
$\{y_i,z_i\}$ are the coordinates of the $i$-th hole.  If $I_{h}(y,z)$
is constant in the scanned spatial region ($r$ is small enough), and
if the hole distribution is $\delta$-correlated, $I(y,z)$ (excluding
the region around $y=0$ and $z=0$) is a
standard speckle with the two-point correlation function $C(\delta
y,\delta z)=\langle I(y,z) I(y+\delta y,z+\delta z)\rangle-\langle I
\rangle^2$ given by
\begin{equation}
 C(\delta y,\delta z)=\langle I\rangle^2\,{\rm sinc}^2\left(\dfrac{L\,\delta y}{\lambda_L f}\right)
{\rm sinc}^2\left(\dfrac{L\,\delta z}{\lambda_L f}\right),
\end{equation}
where $\langle\rangle$ denotes both 
the average over disorder realizations and
over each realization.
This is shown in Fig. \ref{fig-corr} in red dotted line where we have
plotted the rescaled correlation function $c(\delta z)=C(0,\delta
z)/C(0,0)$ (top panel) and the corresponding spectral function
\begin{equation}
S(q)=\int_{-\infty}^{+\infty} e^{-i2\pi q\delta z} c(\delta z)\,d(\delta z),
\end{equation} 
that is the well-known triangular function that goes to zero at
$q=1/\sigma_R$, $\sigma_R=(\lambda_L f)/L$ being the correlation
length (bottom panel). The compact-$q$ support of the speckle is a
result of the finite size of the diffracting plate. These results were
obtained numerically from the random potentials used in the dynamical
simulations of Sec.\ \ref{model}.

The speckle properties are robust to short-distance correlations in
the hole distribution when the correlation range is much smaller than
the plate size \cite{Good07}.  But by introducing hole correlations at
larger distances, $c(\delta z)$ and $S(q)$ can be accordingly
modified.  In particular we consider a hole-dimerized distribution,
where at each hole at position $\{y_i,z_i\}$ corresponds another hole
at position $\{y_i,z_j\}$ with $y_j=y_i$ and $|z_j-z_i|=d$ (see
Fig. \ref{fig-schema}(b)). From a distance the resulting speckle looks
similar to the standard one, but by zooming in the presence of some
order in the grain distribution is clearly observable.  Indeed, the
resulting light pattern, for the case $d<L/2$ where each hole has a
partner at a distance $d$,
\begin{equation}
\begin{split}
I_{RD}(y,z)&\propto\left|\sum_{i=1}^{N_h/2}[ e^{-\frac{2i\pi}{\lambda_L f}(yy_i+zz_i)}+ e^{-\frac{2i\pi}{\lambda_L f}(yy_i+z(z_i+d))}]
\right|^2\\
&=\left|\sum_{i=1}^{N_h/2}e^{-\frac{2i\pi}{\lambda_L f}(yy_i+zz_i)}
\right|^2\{2\cos[\pi zd/(\lambda_L f)]\}^2\\
\label{newspeck}
\end{split}
\end{equation} 
is a product of the standard speckle and a sinusoidal function
with the spatial period $\lambda_L f/d$. 
The correlation function of
such a random-dimer speckle (RD-speckle) corresponds roughly to 
the superposition
of the correlation function for a standard speckle and a sinusoidal
function with a well-defined spatial frequency (see top panel of
Fig. \ref{fig-corr}), that results in a peak at $q=d/(L\sigma_R)$ in
the corresponding spectral function (bottom panel). 
%The correlation
%function of the RD-speckle for $\delta z> \sigma_R$ is very similar to 
%that of the Edwards model with random dimer impurities as obtained in
%\cite{Alamir2012}.
For the case $d\ge L/2$ few holes located at the plate center have no partners.
We have checked that these few holes do not affect the spectral 
function with respect to a case where all holes are dimerized.

Although correlations in the speckle have been previously introduced
by changing the aperture of the diffusive plate or the spatial profile
of the incident beam as proposed in Refs. \cite{Piraud2011,
  Plodzien2011, Piraud2013}, and other strategies have been proposed
in the context of microwave experiments to introduce a non-monotonic
behaviour of the localization efficiency \cite{Izraivel1999,Kuhl08},
the interest of the present RD-speckle lies in the possibility to
control the position of the peak in the spectral function $S(q)$ with
standard experimental techniques. As it will be enlightened in the
following, this property allows to scan, as a function of the energy,
the response of a system to the disorder potential generated by the
light pattern.  Indeed the standard speckle and the RD-speckle can be
used as disorder potentials in an ultracold-atom experiment, the
strength of the potentials being given by
\begin{equation}
{\mathscr V}(y,z)\simeq-I(y,z)\frac{3 \pi c_{\text{light}}^2}{2 \omega_{at}^3}\frac{\Gamma}{\delta}
\end{equation} 
where $c_{\text{light}}$ is the light velocity, $\Gamma$ the linewidth
of the atomic transition, $\delta=\omega_{at}-\omega_L$ the detuning
between the atomic frequency $\omega_{at}$ and the laser frequency
$\omega_L=2\pi c_{\text{light}}/\lambda_L$. Furthermore, as in the
Born approximation the spectral function $S(q)$ is proportional to the
inverse of the localization length, by changing the correlation
properties of the speckle one effectively modifies the localization in
the same manner.

The geometry of the random potential ${\mathscr V}(y,z)$ can be varied
by changing the dimensions of the plate. A 1D random potential
\cite{Billy2008} can be realized for instance by squeezing the
$y$-size of the diffusive plate.  In this way, the transverse size of
the speckle grains can be much larger than the system transverse
size. This is equivalent to considering the 1D potential $V(z)={\mathscr
  V}(y=\bar y,z)$ as it will be done in the following.

\section{Single-particle localization efficiency}
\label{loc} 
With the aim to clarify the effects of a RD-speckle potential in the
Anderson localization frame, we study the propagation of a quantum
particle of mass $m$ along an infinitely long 1D disorder potential
$V(z)$, as described by the time-independent Schr\"{o}dinger equation
\begin{equation}\label{schroedinger}
- \frac{\hbar^2}{2 m} \frac{\partial^2}{\partial z^2} \psi(z) + V(z) \,\psi(z) = E \psi(z),
\end{equation}
$E=\hbar^2k_E^2/2m +\langle V \rangle$ being the particle energy.
The Lyapunov exponent $\gamma(E)$ that coincides with
the inverse of the localization length $\mathcal{L}_{loc}(E)$, 
is given by~\cite{Paladin1987}

\begin{equation}
\gamma(E) = \dfrac{1}{\mathcal{L}_{loc}(E)}=\lim_{|z| \rightarrow \infty} \frac{1}{|z|} \left\langle \ln \left( \frac{k^2_E \psi^2(z) + \psi'^2 (z)}{k^2_E \psi^2(0) + \psi'^2 (0)} \right) \right\rangle .
\label{eq-gamma}
\end{equation}
We compute numerically $\gamma(E)$ by discretizing the Schr\"{o}dinger
equation on a spatial grid and writing the equation in a matrix form:
\begin{equation}\label{eq-discret}
\begin{pmatrix}
& \psi_{n+1} \\
& \psi_n
\end{pmatrix}
= T_n
\begin{pmatrix}
& \psi_n \\
& \psi_{n-1} 
\end{pmatrix} 
\end{equation}
where $\psi_n$ is the wave function at the grid position $n$ and $T_n$
is a so-called transfer matrix. The final wave vector
$(\psi_{n+1},\psi_n)^T$ is found by plugging in an initial vector
$(\psi_{1},\psi_0)^T$ and solving recursively Eq.~\eqref{eq-discret}.
We used as initial wave $(\psi_0, \partial_z \psi_0 ) = (1, k/ \tan
\theta )$ where $\theta \in [0, 2 \pi]$ is a random
angle~\cite{piraudthesis} and we exploited the Numerov algorithm
\cite{Chow1972} to write Eq. (\ref{eq-discret}) at each spatial point.
We propagated the wave over a grid of $4 \times 10^6$ points with step
size 0.1$\sigma_R$, such that the details of the speckle function are
taken into account, and averaged over $10^4$ realizations.

\begin{figure}
\includegraphics[width=1\linewidth]{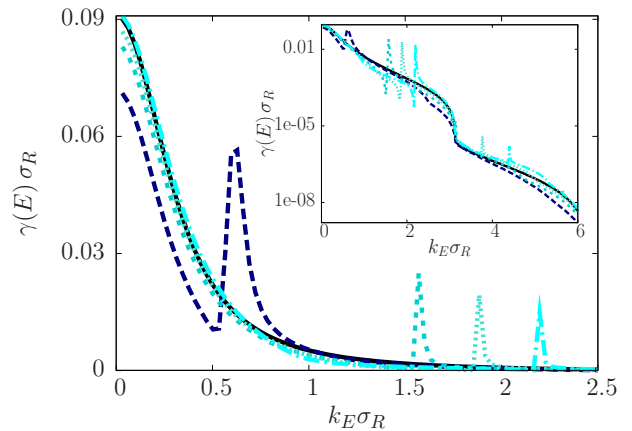}
\caption{\label{fig-loc}(Color online) Lyapunov exponent $\gamma(E)$ in unit of
  $1/\sigma_R$ as a function of $k_E\sigma_R$ for a standard speckle
  potential (continuous black line) and for RD-speckle potentials with
 $d=0.4, 1, 1.2$ and $1.4$ cm (colored dashed lines) from left to right.
The inset shows the same curve but in log-scale, over a wider $k_E$ range.
%(dashed dark-blue line), $d=1$ cm (short-dashed
%  dark-turquoise line), $d=1.2$ cm (dotted turquoise line) and $d=1.4$
%  cm (dot-dashed cyan line)
}
\end{figure}

In Fig. \ref{fig-loc} we show the behavior of $\gamma(E)$ as a
function of $k_E$ for the case of a $^{87}$Rb atom ($\lambda_{at}=2\pi
c_{\text{light}}/\omega_{at}=780$ nm and $\Gamma=2 \pi \times 6.065$
MHz) subjected to a disorder potential of strength
$V_{dis}=\sqrt{\langle V^2\rangle} = 0.117\, \hbar^2 / m \sigma_R^2$,
generated by a laser of wavelength $\lambda_L = 532$ nm.  We compare
the case of a standard speckle (continuous black line) with that of a
RD-speckle for different values of $d$.  As shown for the spectral
function $S(q)$ (Fig. \ref{fig-corr}), in the RD case a peak appears
whose position depends linearly on $d$ (dashed, dotted and dot-dashed
lines in Fig. \ref{fig-loc}).  Thus the RD-speckle potential allows to
achieve and control the localization of high-energy atoms.

Except for the main peak, $\gamma(E)$ of the RD-speckle has the same
trend as the conventional speckle displaying the effective mobility
edge at $k_E=\pi/\sigma_R$ \cite{Luga09}.  This was predictable from
the calculation of the spectral function $S(q)$ (see bottom panel of
Fig. \ref{fig-corr}) since the Lyapunov coefficient evaluated in the
Born approximation is proportional to $S(2q)$ and $k_E=2\pi q$.  The
overall behavior of the RD-speckle is clearly observable in the inset
of Fig. \ref{fig-loc} where we have plotted $\gamma(E)$ using a
logarithmic scale and over a larger range of $k_E$. In particular, we
observe several low amplitude revivals at higher energies located at
integer multiples of the position of the main peak.

\section{Dynamics of a quasi-one dimensional BEC}
\label{model}
We study the dynamics of a system of $N=10^5$ Bose-Einstein 
condensed $^{87}$ Rb atoms of mass $m$ subject
to a static cigar-shaped
harmonic trap and a time-dependent random potential: 
\begin{equation}
\label{potential1}
U(\textbf{r},t) = \frac{1}{2} m \omega^2_\perp (x^2 + y^2) +
\frac{1}{2} m \omega^2_z z^2 + V(z,t) 
\end{equation} 
with $\omega_{\perp}=2 \pi \times 235.8$ Hz and $\omega_{z} = 2 \pi
\times 22.2$ Hz the trapping frequencies in the perpendicular and
longitudinal directions, respectively.  The last time-dependent term
in (\ref{potential1}) corresponds to a random potential that is fixed
in the moving frame $z'=z-vt$, ${\bf v}=v\hat e_z$ being the drift
velocity.  The random potential is generated by the procedure
illustrated in Sec. \ref{speckle}.

Under cigar-shaped trap geometry, the full 3D equation of motion for
the BEC wavefunction $\psi(\textbf{r},t)$ can be reduced to the
effective 1D time-dependent nonpolynomial nonlinear Schr\"odinger
equation (NPSE)~\cite{salasnich02}
\begin{equation}
\begin{split}
i \hbar \frac{\partial}{\partial t} f &= \left[ - \frac{\hbar^2}{2 m} \frac{\partial^2}{\partial z^2} + \frac{1}{2} m \omega^2_z z^2 + V(z,t)
\right.\\
&\left.+ \hbar \omega_{\perp} 
\frac{1 + 3 a_s N |f|^2 }{\sqrt{1 + 2 a_s N |f|^2}} \right] f.
\label{NPSE}
\end{split}
\end{equation}
with $a_s$ being the {\it s}-wave scattering length that we set at $80 a_B$ and
$a_B$ being the Bohr radius. 
To obtain Eq. (\ref{NPSE}) we set 
\begin{equation}
\psi(\textbf{r},t) = f(z,t) \phi(\textbf{r},t)=f(z,t)
\frac{e^{-(x^2+y^2) / 2 \sigma^2(z,t)} }{\sqrt{\pi} \sigma(z,t)} 
\end{equation}
where the transverse part $\phi(\textbf{r},t)$ is modeled by a
Gaussian function with variance $\sigma(z,t)$.
Within the assumption that this variance varies slowly 
as functions of $z$ and $t$, $\sigma(z,t)$ is given by
\begin{equation}
\sigma^2 (z,t) = \ell_0^2\sqrt{1 + 2 a_s N |f(z,t)|^2},
\end{equation}
where $\ell_0 = \sqrt{\hbar / (m \omega_\perp)}$ is the oscillator 
length in the transverse direction.
The 3D density profile is then	 
\begin{equation}
\rho (\textbf{r})  = \tilde{\rho} (z) \frac{e^{-(x^2+y^2) / \sigma^2}}{\pi \sigma^2}, 
\label{thur07071} 
\end{equation}
with $\tilde{\rho} (z) = |f|^2$ the integrated 1D density. 

The NPSE is numerically solved using a split-step method and spatial
Fast Fourier transforms (FFT). First we compute the equilibrium
density profile in the presence of a static disorder potential. Then,
we switch on the drift velocity $v$ and compute the time evolution of
the condensate wavefunction $f(z,t)$.

\section{Quantum versus classical transport}
\label{results}
The scheme of the proposed experiment is the following. The disorder
potential is pulled through the BEC with a velocity $v$ over a
distance $L^*$.  We measure the center-of-mass shift $z_{cm}$ and we
identify the ratio of localized atoms $N_{loc}/N$ with the ratio
$z_{cm}/L^*$, indeed if the whole BEC is insensible to the disorder
potential then $z_{cm}=0$, while if the whole BEC is stuck on the
disorder potential then $z_{cm}=L^*$ \cite{Alamir2012}.

\begin{figure}
\includegraphics[width=1.\linewidth]{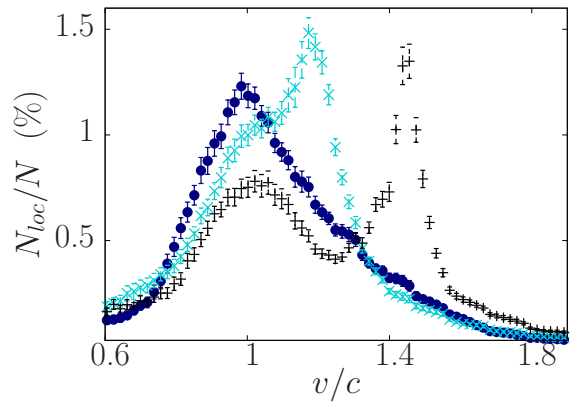}
\caption{(Color online)\label{fig-d} Localized BEC
  fraction as a function of $v/c$ for the case of a RD-speckle
with $d=0.4$ cm (blue circles), $d=0.8$ cm (turquoise crosses) 
and $d=1.2$ cm (black plus signs).}
\end{figure}

Since we expect to observe localization for $v\gtrsim c$,
$c=\sqrt{\mu/2m}$ being the 1D speed of sound
\cite{leboeuf09,Albert2010,Alamir2012} with $\mu$ the chemical
potential, we tune the position of the localization peak in this
region and choose the value of the dimer length $d$ to enhance the
interference effects.  With this purpose, we study the localized BEC
fraction $N_{loc}/N$ as a function of $v/c$ for the case of a
RD-speckle of potential strength $V_{dis}=0.05\,\hbar\omega_\perp$ and
different values of $d$, $d=0.4$ cm (blue circles), $d=0.8$ cm
(turquoise crosses) and $d=1.2$ cm (black plus signs).  This is
shown in Fig. \ref{fig-d} where we can observe that the best resolved
peak corresponds to $d=1.2$ cm.  By identifying the drift velocity $v$
with $\hbar k_E/m$, this corresponds to a $\gamma(E)$ peak at $v\simeq
1.1 c$.
Here and in the following we average over 30 configurations; in all
simulations we fix $L^*\simeq 56\sigma_R$, value that corresponds to
$\sim 1.4$ times ${\mathcal L}_{loc}$ evaluated at the peak position
of the case $d=1.2$ cm, which is the $d$ value that we set from now
on.

\begin{figure}
\includegraphics[width=1.\linewidth]{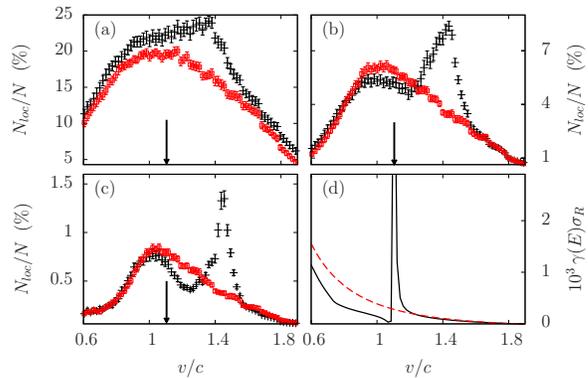}
\caption{(Color online)\label{fig-Nloc} Panels (a-c): localized BEC
  fraction as a function of $v/c$ for the case of a standard speckle
  (red square) and of a RD-speckle (black crosses).  (a)
  $V_{dis}=0.39\,\hbar\omega_\perp$, (b)
  $V_{dis}=0.16\,\hbar\omega_\perp$ and (c)
  $V_{dis}=0.05\,\hbar\omega_\perp$. The vertical arrows indicate the
  position of the $\gamma(E)$ peak. Panel (d): $\gamma(E)$ in units
  $1/\sigma_R$ as a function of $v/c$ with $v=\hbar k_E/m$ for the
  standard speckle (dashed lines) and RD speckle (solid lines). All
  calculations correspond to $d=1.2$ cm.}
\end{figure}

The localized BEC fraction as a function of $v/c$ is shown in Fig.\
\ref{fig-Nloc} for different values of the potential strength
$V_{dis}$ for the case of a standard speckle (red squares) and a
RD-speckle (black crosses).
We observe that at large values of $V_{dis}$, the behaviors of
$N_{loc}/N$ of the standard and the RD speckles are quite similar
(panel (a) of Fig. \ref{fig-Nloc}).  By lowering $V_{dis}$, the global
localization efficiency of the disorder potential decreases but a peak
appears at $v/c\simeq 1.4$ for the case of a RD-speckle (panels (b)
and (c) of Fig. \ref{fig-Nloc}), as already outlined in
Fig. \ref{fig-d}. Moreover this peak is preceded by a strong
inhibition of the localization with respect to the standard speckle in
agreement with the behaviour of $\gamma(E)$ (panel (d) in
Fig. \ref{fig-Nloc}) where the curves for the standard and the RD
speckles intersect before the peak.

Although the peak in $N_{loc}/N(v)$ can be
attributed to the interference effects giving rise to the one of
$\gamma(E)$, its position is shifted and the shape broader. Indeed the
calculation of $\gamma(E)$ is done in the single-particle
approximation for a steady potential, while the BEC is an
inhomogeneous and interacting many-particle system. The system is then
continuously disturbed by pulling the disorder potential; thus many
factors may contribute to the shift and the broadening of the peak.
We can also observe that the peak widens by increasing the disorder
strength $V_{dis}$.

\begin{figure}
\includegraphics[width=0.75\linewidth]{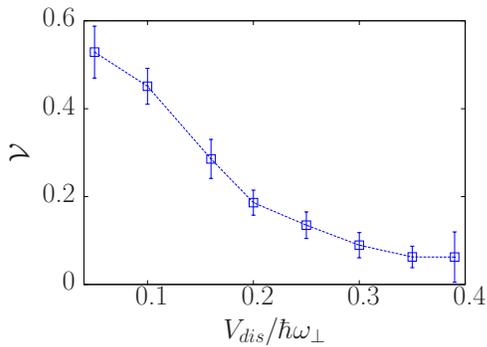}
\caption{(Color online)\label{fig-vis} Visibility $\mathcal{V}$ 
of the peak of the function $N_{loc}/N$ as a function of $V_{dis}$ in unit of
$\hbar\omega_\perp$. The line is a guide to the eye.}
\end{figure}

The behavior of the $N_{loc}/N$ peak for the RD-speckle as a function
of $V_{dis}$ is shown in Fig. \ref{fig-vis}. In this figure we have
plotted the visibility $\mathcal{V}$ of the peak, defined as
\begin{equation}
  \mathcal{V}=\dfrac{\left(\dfrac{N_{loc}}{N}\right)_{\max}-\left(\dfrac{N_{loc}}{N}\right)_{\min}}{\left(\dfrac{N_{loc}}{N}\right)_{\max}+\left(\dfrac{N_{loc}}{N}\right)_{\min}}
\end{equation}
where $\left(N_{loc}/N\right)_{\max}$ and
$\left(N_{loc}/N\right)_{\min}$ are respectively the peak and the
hollow preceding the peak of the function
$\left(N_{loc}/N\right)(v/c)$.
For all values of $V_{dis}$ considered in Fig. \ref{fig-vis}, the
chemical potential $\mu$ is of the order of $7.8 \hbar\omega_{\perp}$, and
thus the drift kinetic energy at the peak location ($v\simeq 1.4\,c$)
is of the order of $2\mu\simeq 15.6\,\hbar\omega_{\perp}$ a quite large value
with respect to the potential strengths $V_{dis}$ considered in this
work.  However in the standard speckle, as well as in the RD-one, the
probability for high-intensity grains is not vanishing and the BEC can
be trapped by few potential wells to quite low values of $V_{dis}$ as
it happens for the case $V_{dis}=0.39\,\hbar\omega_\perp$ (panel (a)
of Fig. \ref{fig-Nloc}).  Moreover, because of the sinusoidal function
[see Eq. (\ref{newspeck})] that modulates the standard speckle on a
smaller scale with respect the size grains, the probability to have
very high grains is larger for the RD-speckle than for the standard
speckle. This explains the fact that at
$V_{dis}=0.39\,\hbar\omega_\perp$, the localization efficiency of the
RD-speckle is larger than that of the standard speckle over the whole
$v/c$ range.  In order to observe interference effects in the
localization dynamics, one needs to reduce $V_{dis}$ further so that
to decrease the probability to have speckle grains over a given
threshold. Indeed the peak in the $N_{loc}/N$ function becomes clearly
visible ($\mathcal{V}\ge 0.2$) at $V_{dis}\le 0.16\hbar\omega_\perp$.

\begin{figure}
\includegraphics[width=1.\linewidth]{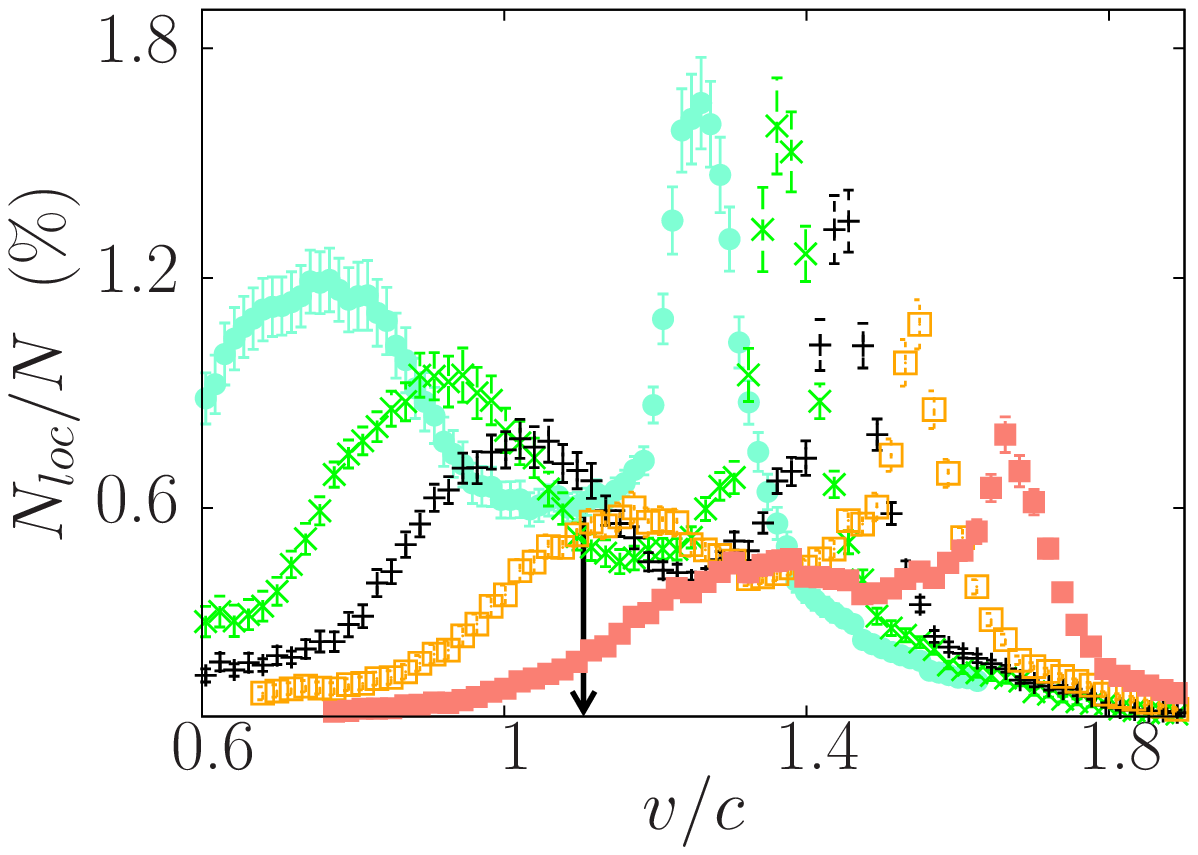}
\includegraphics[width=1.\linewidth]{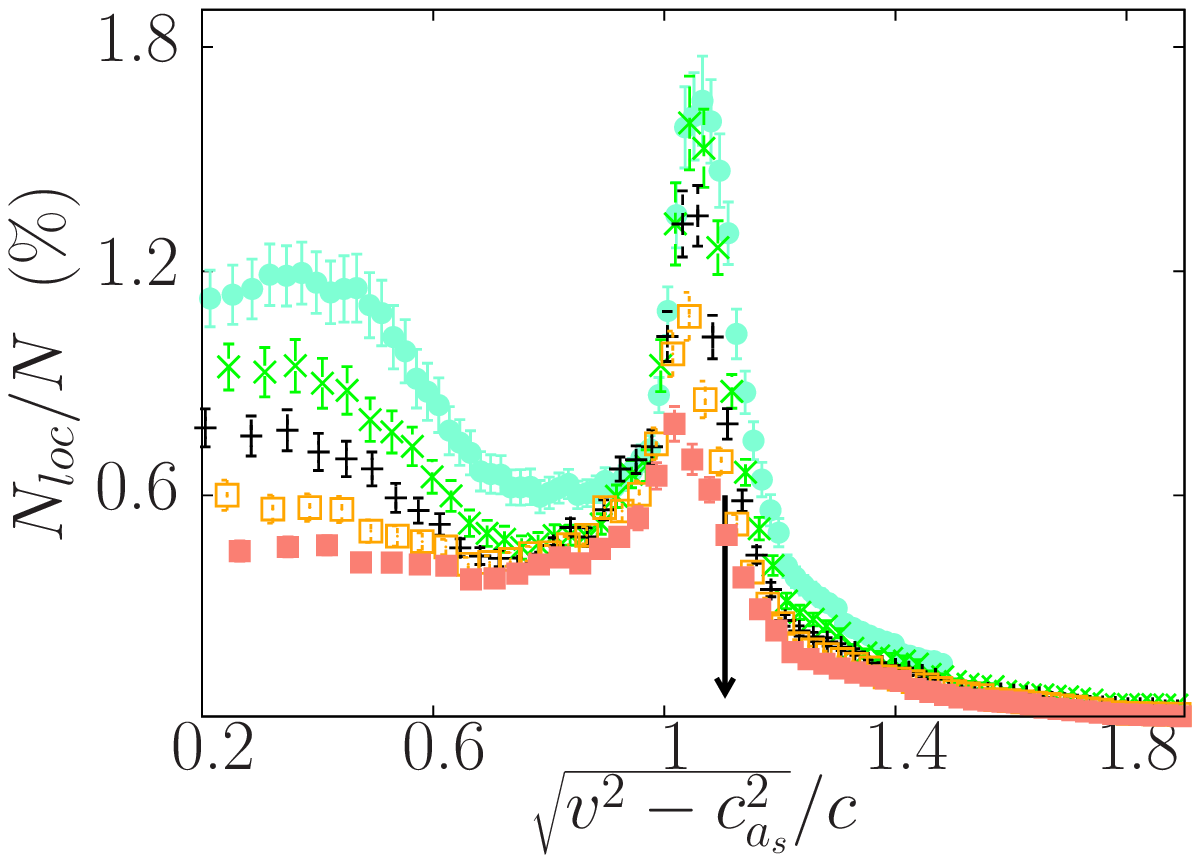}
\caption{(Color online)\label{fig-int} Top panel: Localized BEC fraction as a
  function of $v/c$ ($c$ being the sound velocity for the case $a_s=80
  a_B$) for the case of a RD-speckle with $d=1.2$ cm and
  $V_{dis}=0.05\,\hbar\omega_\perp$. The different symbols correspond
  to $a_s=10 a_B$ (aquamarine filled circles), $40a_B$ (green 
  crosses), $80 a_B$ (black plus signs), $160 a_B$ (orange empty squares)
  and $320 a_B$ (pink filled squares). The vertical arrows indicate
  the position of the $\gamma(E)$ peak. For each value of $a_s$ we
  average over 30 configurations. Bottom panel: The same as in the top panel, but the localized BEC fraction is shown as a function 
of $\sqrt{v^2-c_{a_{s}}^2}/c$, with $c_{a_{s}}$ being the sound velocity for the corresponding $a_s$.}
\end{figure}

For a better understanding of the role of the interactions on the peak
distorsion and, more generally, on the localization efficiency, we
vary the scattering length $a_s$ from 10 to 320 $a_B$ for fixed
potential strength $V_{dis}=0.05\,\hbar\omega_\perp$. Let us remark
that this is just a conceptual experiment since the $^{87}$Rb
scattering length cannot be tuned by exploiting Feshbach resonances.
The results are shown in the top panel of Fig. \ref{fig-int}, where we
have drawn the localization efficiency for different interaction
strengths as a function of $v/c$, $c$ being the sound velocity for the
case $a_s=80 a_B$ in order to fix the same velocity scale for all
curves.
The first observation is that the larger the value of the interaction
is, the greater the shift of the position of the peak is.  Indeed, in
the presence of interactions, what really matters is the available
kinetic energy $\frac{1}{2}mv^2-\frac{1}{2}mc_{a_{s}}^2$
\cite{Leboeuf2001}, where $c_{a_s}$ is the sound velocity
corresponding at the scattering length $a_{s}$. Actually, if we plot
$N_{loc}/N$ as a function of $\sqrt{v^2-c_{a_{s}}^2}/c$ (bottom panel
of Fig. \ref{fig-int}), all the peaks collapse at the same point, very
close to the position of the $\gamma(E)$ peak.  The second observation
is that, by increasing interactions, the localization efficiency of
the RD-speckle decreases overall in the $v$ space. By increasing the
interactions we increase the robustness of the superfluidity, and the
disorder potential becomes less and less efficient to localize the
atoms. Indeed, if we consider, for example, the case $a_s=320 a_B$ the
position of the $\gamma(E)$ peak in units of $v/c_{320}$, $c_{320}$
being the corresponding sound velocity, we find $\sim 0.8 v/c_{320}$
and thus is not the best value of $v$ for this scattering
length. Therefore, to increase the efficiency of our potential at
large interaction strengths, we should increase the value of $d$.

Finally we would like to stress that the measure of $N_{loc}/N$ by
means of the center-of-mass shift is really a good observable to
detect localization in a trapped system, both for the standard speckle
and for our new proposed speckle. Indeed its statistical distribution
function displays the expected behavior in the presence of
localization.
\begin{figure}
\includegraphics[width=0.98\linewidth]{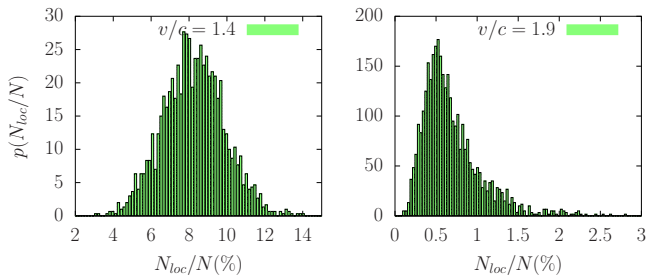}
\caption{(Color online)\label{fig-isto} Statistical distribution
  $p(N_{loc}/N)$ for the RD-speckle with $V_{dis}=0.16\hbar\omega_\perp$,
  $d=1.2$ cm and for two values of $v$.}
\end{figure}
This is shown in Fig. \ref{fig-isto} where we compare the statistical
distribution $p(N_{loc}/N)$ for a RD speckle at two drift
velocities. We observe that at the velocity corresponding to the
localization maxima (left panel) the statistical distribution becomes
rather symmetric as opposed to the shape for the larger $v$. This is
in analogy to the behaviour of the transmission in a non-interacting
homogeneous system \cite{Muller2010}, where one could expect a
transition from a decreasing exponential distribution at vanishingly
low localization ratio to a log-normal distribution when localization
dominates. A complete analysis on this matter requires a thorough
study entailing a systematic calculation of the statistical properties
of the potential for many realizations, drift velocities, distances
$d$, etc and therefore is left for future investigations.

\section{Conclusions}
\label{concl}
We studied the dragging of a Bose-Einstein condensate of $^{87}$Rb
atoms confined in cigar-shaped traps in the presence of a correlated
speckle potential. By constructing a speckle out of randomly
distributed dimerized holes, we are able to select a non-vanishing
energy value that maximizes the localization efficiency and thus to
localize higher-energy atoms.  Our approach can be implemented by
spatial light modulator devices available as standard experimental
equipment.  By numerically solving the dynamics of the condensate
subjected to an underlying disorder potential moving at constant
speed, we have shown the efficacity and versatility of such a
potential.  By analyzing the center-of-mass displacement, we find that
correlations enhance the localization by a factor of 2-3 with respect
to standard speckle. The magnitude of this effect is very sensitive to
the interaction strength and the amplitude of the disorder. Indeed a
strong disorder inhibits interference thwarting the presence of
correlations in the condensate dynamics.

\begin{acknowledgments}
  This work was supported by CNRS PICS grant No. 05922. P. C. acknowledges
  support  ANPCyT 2008-0682 and PIP 0546 from CONICET. The authors aknowledge
M. Albert, C. Miniatura, G. Modugno, N. Pavloff and 
L. Tessieri for useful discussions.
\end{acknowledgments}

 %\bibliography{biblio}

\begin{thebibliography}{34}
\expandafter\ifx\csname natexlab\endcsname\relax\def\natexlab#1{#1}\fi
\expandafter\ifx\csname bibnamefont\endcsname\relax
  \def\bibnamefont#1{#1}\fi
\expandafter\ifx\csname bibfnamefont\endcsname\relax
  \def\bibfnamefont#1{#1}\fi
\expandafter\ifx\csname citenamefont\endcsname\relax
  \def\citenamefont#1{#1}\fi
\expandafter\ifx\csname url\endcsname\relax
  \def\url#1{\texttt{#1}}\fi
\expandafter\ifx\csname urlprefix\endcsname\relax\def\urlprefix{URL }\fi
\providecommand{\bibinfo}[2]{#2}
\providecommand{\eprint}[2][]{\url{#2}}

\bibitem[{\citenamefont{Anderson}(1958)}]{Ande58}
\bibinfo{author}{\bibfnamefont{P.~W.} \bibnamefont{Anderson}},
  \bibinfo{journal}{Phys. Rev.} \textbf{\bibinfo{volume}{109}},
  \bibinfo{pages}{1492} (\bibinfo{year}{1958}).

\bibitem[{\citenamefont{Anderson}(1985)}]{Ande85}
\bibinfo{author}{\bibfnamefont{P.~W.} \bibnamefont{Anderson}},
  \bibinfo{journal}{Philosophical Magazine Part B}
  \textbf{\bibinfo{volume}{52}}, \bibinfo{pages}{505} (\bibinfo{year}{1985}).

\bibitem[{\citenamefont{Eiermann et~al.}(2004)\citenamefont{Eiermann, Anker,
  Albiez, Taglieber, Treutlein, Marzlin, and Oberthaler}}]{Eiermann2004}
\bibinfo{author}{\bibfnamefont{B.}~\bibnamefont{Eiermann}},
  \bibinfo{author}{\bibfnamefont{T.}~\bibnamefont{Anker}},
  \bibinfo{author}{\bibfnamefont{M.}~\bibnamefont{Albiez}},
  \bibinfo{author}{\bibfnamefont{M.}~\bibnamefont{Taglieber}},
  \bibinfo{author}{\bibfnamefont{P.}~\bibnamefont{Treutlein}},
  \bibinfo{author}{\bibfnamefont{K.-P.} \bibnamefont{Marzlin}},
  \bibnamefont{and} \bibinfo{author}{\bibfnamefont{M.~K.}
  \bibnamefont{Oberthaler}}, \bibinfo{journal}{Phys. Rev. Lett.}
  \textbf{\bibinfo{volume}{92}}, \bibinfo{pages}{230401}
  (\bibinfo{year}{2004}),
  \urlprefix\url{http://link.aps.org/doi/10.1103/PhysRevLett.92.230401}.

\bibitem[{\citenamefont{Fisher et~al.}(1989)\citenamefont{Fisher, Weichman,
  Grinstein, and Fisher}}]{Fisher1989}
\bibinfo{author}{\bibfnamefont{M.~P.~A.} \bibnamefont{Fisher}},
  \bibinfo{author}{\bibfnamefont{P.~B.} \bibnamefont{Weichman}},
  \bibinfo{author}{\bibfnamefont{G.}~\bibnamefont{Grinstein}},
  \bibnamefont{and} \bibinfo{author}{\bibfnamefont{D.~S.}
  \bibnamefont{Fisher}}, \bibinfo{journal}{Phys. Rev. B}
  \textbf{\bibinfo{volume}{40}}, \bibinfo{pages}{546} (\bibinfo{year}{1989}),
  \urlprefix\url{http://link.aps.org/doi/10.1103/PhysRevB.40.546}.

\bibitem[{\citenamefont{Scalettar et~al.}(1991)\citenamefont{Scalettar,
  Batrouni, and Zimanyi}}]{Scalettar1991}
\bibinfo{author}{\bibfnamefont{R.~T.} \bibnamefont{Scalettar}},
  \bibinfo{author}{\bibfnamefont{G.~G.} \bibnamefont{Batrouni}},
  \bibnamefont{and} \bibinfo{author}{\bibfnamefont{G.~T.}
  \bibnamefont{Zimanyi}}, \bibinfo{journal}{Phys. Rev. Lett.}
  \textbf{\bibinfo{volume}{66}}, \bibinfo{pages}{3144} (\bibinfo{year}{1991}),
  \urlprefix\url{http://link.aps.org/doi/10.1103/PhysRevLett.66.3144}.

\bibitem[{\citenamefont{Ristivojevic et~al.}(2012)\citenamefont{Ristivojevic,
  Petkovi\ifmmode~\acute{c}\else \'{c}\fi{}, Le~Doussal, and
  Giamarchi}}]{Ristivojevic2012}
\bibinfo{author}{\bibfnamefont{Z.}~\bibnamefont{Ristivojevic}},
  \bibinfo{author}{\bibfnamefont{A.}~\bibnamefont{Petkovi\ifmmode~\acute{c}\else
  \'{c}\fi{}}}, \bibinfo{author}{\bibfnamefont{P.}~\bibnamefont{Le~Doussal}},
  \bibnamefont{and}
  \bibinfo{author}{\bibfnamefont{T.}~\bibnamefont{Giamarchi}},
  \bibinfo{journal}{Phys. Rev. Lett.} \textbf{\bibinfo{volume}{109}},
  \bibinfo{pages}{026402} (\bibinfo{year}{2012}),
  \urlprefix\url{http://link.aps.org/doi/10.1103/PhysRevLett.109.026402}.

\bibitem[{\citenamefont{Pilati et~al.}(2009)\citenamefont{Pilati, Giorgini, and
  Prokof'ev}}]{Pilati2009}
\bibinfo{author}{\bibfnamefont{S.}~\bibnamefont{Pilati}},
  \bibinfo{author}{\bibfnamefont{S.}~\bibnamefont{Giorgini}}, \bibnamefont{and}
  \bibinfo{author}{\bibfnamefont{N.}~\bibnamefont{Prokof'ev}},
  \bibinfo{journal}{Phys. Rev. Lett.} \textbf{\bibinfo{volume}{102}},
  \bibinfo{pages}{150402} (\bibinfo{year}{2009}),
  \urlprefix\url{http://link.aps.org/doi/10.1103/PhysRevLett.102.150402}.

\bibitem[{\citenamefont{Allard et~al.}(2012)\citenamefont{Allard, Plisson,
  Holzmann, Salomon, Aspect, Bouyer, and Bourdel}}]{Allard2012}
\bibinfo{author}{\bibfnamefont{B.}~\bibnamefont{Allard}},
  \bibinfo{author}{\bibfnamefont{T.}~\bibnamefont{Plisson}},
  \bibinfo{author}{\bibfnamefont{M.}~\bibnamefont{Holzmann}},
  \bibinfo{author}{\bibfnamefont{G.}~\bibnamefont{Salomon}},
  \bibinfo{author}{\bibfnamefont{A.}~\bibnamefont{Aspect}},
  \bibinfo{author}{\bibfnamefont{P.}~\bibnamefont{Bouyer}}, \bibnamefont{and}
  \bibinfo{author}{\bibfnamefont{T.}~\bibnamefont{Bourdel}},
  \bibinfo{journal}{Phys. Rev. A} \textbf{\bibinfo{volume}{85}},
  \bibinfo{pages}{033602} (\bibinfo{year}{2012}),
  \urlprefix\url{http://link.aps.org/doi/10.1103/PhysRevA.85.033602}.

\bibitem[{\citenamefont{Onofrio et~al.}(2000)\citenamefont{Onofrio, Raman,
  Vogels, Abo-Shaeer, Chikkatur, and Ketterle}}]{Onofrio2000}
\bibinfo{author}{\bibfnamefont{R.}~\bibnamefont{Onofrio}},
  \bibinfo{author}{\bibfnamefont{C.}~\bibnamefont{Raman}},
  \bibinfo{author}{\bibfnamefont{J.~M.} \bibnamefont{Vogels}},
  \bibinfo{author}{\bibfnamefont{J.~R.} \bibnamefont{Abo-Shaeer}},
  \bibinfo{author}{\bibfnamefont{A.~P.} \bibnamefont{Chikkatur}},
  \bibnamefont{and} \bibinfo{author}{\bibfnamefont{W.}~\bibnamefont{Ketterle}},
  \bibinfo{journal}{Phys. Rev. Lett.} \textbf{\bibinfo{volume}{85}},
  \bibinfo{pages}{2228} (\bibinfo{year}{2000}),
  \urlprefix\url{http://link.aps.org/doi/10.1103/PhysRevLett.85.2228}.

\bibitem[{\citenamefont{Astrakharchik and
  Pitaevskii}(2004)}]{Astrakharchik2004}
\bibinfo{author}{\bibfnamefont{G.~E.} \bibnamefont{Astrakharchik}}
  \bibnamefont{and} \bibinfo{author}{\bibfnamefont{L.~P.}
  \bibnamefont{Pitaevskii}}, \bibinfo{journal}{Phys. Rev. A}
  \textbf{\bibinfo{volume}{70}}, \bibinfo{pages}{013608}
  (\bibinfo{year}{2004}),
  \urlprefix\url{http://link.aps.org/doi/10.1103/PhysRevA.70.013608}.

\bibitem[{\citenamefont{Ianeselle et~al.}(2006)\citenamefont{Ianeselle,
  Menotti, and Smerzi}}]{Ianeselli2006}
\bibinfo{author}{\bibfnamefont{S.}~\bibnamefont{Ianeselle}},
  \bibinfo{author}{\bibfnamefont{C.}~\bibnamefont{Menotti}}, \bibnamefont{and}
  \bibinfo{author}{\bibfnamefont{A.}~\bibnamefont{Smerzi}},
  \bibinfo{journal}{J. Phys. B} \textbf{\bibinfo{volume}{39}},
  \bibinfo{pages}{S135} (\bibinfo{year}{2006}).

\bibitem[{\citenamefont{Paul et~al.}(2007)\citenamefont{Paul, Schlagheck,
  Leboeuf, and Pavloff}}]{leboeuf07}
\bibinfo{author}{\bibfnamefont{T.}~\bibnamefont{Paul}},
  \bibinfo{author}{\bibfnamefont{P.}~\bibnamefont{Schlagheck}},
  \bibinfo{author}{\bibfnamefont{P.}~\bibnamefont{Leboeuf}}, \bibnamefont{and}
  \bibinfo{author}{\bibfnamefont{N.}~\bibnamefont{Pavloff}},
  \bibinfo{journal}{Phys. Rev. Lett.} \textbf{\bibinfo{volume}{98}},
  \bibinfo{pages}{210602} (\bibinfo{year}{2007}).

\bibitem[{\citenamefont{Paul et~al.}(2009)\citenamefont{Paul, Albert,
  Schlagheck, Leboeuf, and Pavloff}}]{leboeuf09}
\bibinfo{author}{\bibfnamefont{T.}~\bibnamefont{Paul}},
  \bibinfo{author}{\bibfnamefont{M.}~\bibnamefont{Albert}},
  \bibinfo{author}{\bibfnamefont{P.}~\bibnamefont{Schlagheck}},
  \bibinfo{author}{\bibfnamefont{P.}~\bibnamefont{Leboeuf}}, \bibnamefont{and}
  \bibinfo{author}{\bibfnamefont{N.}~\bibnamefont{Pavloff}},
  \bibinfo{journal}{Phys. Rev. A} \textbf{\bibinfo{volume}{80}},
  \bibinfo{pages}{033615} (\bibinfo{year}{2009}).

\bibitem[{\citenamefont{Alamir et~al.}(2012)\citenamefont{Alamir, Capuzzi, and
  Vignolo}}]{Alamir2012}
\bibinfo{author}{\bibfnamefont{A.}~\bibnamefont{Alamir}},
  \bibinfo{author}{\bibfnamefont{P.}~\bibnamefont{Capuzzi}}, \bibnamefont{and}
  \bibinfo{author}{\bibfnamefont{P.}~\bibnamefont{Vignolo}},
  \bibinfo{journal}{Phys. Rev. A} \textbf{\bibinfo{volume}{86}},
  \bibinfo{pages}{063637} (\bibinfo{year}{2012}),
  \urlprefix\url{http://link.aps.org/doi/10.1103/PhysRevA.86.063637}.

\bibitem[{\citenamefont{Alamir et~al.}(2013)\citenamefont{Alamir, Capuzzi, and
  Vignolo}}]{Alamir2013}
\bibinfo{author}{\bibfnamefont{A.}~\bibnamefont{Alamir}},
  \bibinfo{author}{\bibfnamefont{P.}~\bibnamefont{Capuzzi}}, \bibnamefont{and}
  \bibinfo{author}{\bibfnamefont{P.}~\bibnamefont{Vignolo}},
  \bibinfo{journal}{Eur. Phys. J. Special Topics}
  \textbf{\bibinfo{volume}{217}}, \bibinfo{pages}{63} (\bibinfo{year}{2013}).

\bibitem[{\citenamefont{Cl\'{e}ment et~al.}(2005)\citenamefont{Cl\'{e}ment,
  Var\'{o}n, Hugbart, Retter, Bouyer, Sanchez-Palencia, Gangardt, Shlyapnikov,
  and Aspect}}]{Clem05}
\bibinfo{author}{\bibfnamefont{D.}~\bibnamefont{Cl\'{e}ment}},
  \bibinfo{author}{\bibfnamefont{A.~F.} \bibnamefont{Var\'{o}n}},
  \bibinfo{author}{\bibfnamefont{M.}~\bibnamefont{Hugbart}},
  \bibinfo{author}{\bibfnamefont{J.~A.} \bibnamefont{Retter}},
  \bibinfo{author}{\bibfnamefont{P.}~\bibnamefont{Bouyer}},
  \bibinfo{author}{\bibfnamefont{L.}~\bibnamefont{Sanchez-Palencia}},
  \bibinfo{author}{\bibfnamefont{D.~M.} \bibnamefont{Gangardt}},
  \bibinfo{author}{\bibfnamefont{G.~V.} \bibnamefont{Shlyapnikov}},
  \bibnamefont{and} \bibinfo{author}{\bibfnamefont{A.}~\bibnamefont{Aspect}},
  \bibinfo{journal}{Physical Review Letters} \textbf{\bibinfo{volume}{95}},
  \bibinfo{eid}{170409} (pages~\bibinfo{numpages}{4}) (\bibinfo{year}{2005}),
  \urlprefix\url{http://link.aps.org/abstract/PRL/v95/e170409}.

\bibitem[{\citenamefont{Billy et~al.}(2008)\citenamefont{Billy, Josse, Zuo,
  Bernard, Hambrecht, Lugan, Cl\'{e}ment, Sanchez-Palencia, Bouyer, and
  Aspect}}]{Billy2008}
\bibinfo{author}{\bibfnamefont{J.}~\bibnamefont{Billy}},
  \bibinfo{author}{\bibfnamefont{V.}~\bibnamefont{Josse}},
  \bibinfo{author}{\bibfnamefont{Z.}~\bibnamefont{Zuo}},
  \bibinfo{author}{\bibfnamefont{A.}~\bibnamefont{Bernard}},
  \bibinfo{author}{\bibfnamefont{B.}~\bibnamefont{Hambrecht}},
  \bibinfo{author}{\bibfnamefont{P.}~\bibnamefont{Lugan}},
  \bibinfo{author}{\bibfnamefont{D.}~\bibnamefont{Cl\'{e}ment}},
  \bibinfo{author}{\bibfnamefont{L.}~\bibnamefont{Sanchez-Palencia}},
  \bibinfo{author}{\bibfnamefont{P.}~\bibnamefont{Bouyer}}, \bibnamefont{and}
  \bibinfo{author}{\bibfnamefont{A.}~\bibnamefont{Aspect}},
  \bibinfo{journal}{Nature} \textbf{\bibinfo{volume}{453}}, \bibinfo{pages}{891
  } (\bibinfo{year}{2008}).

\bibitem[{\citenamefont{Kondov et~al.}(2011)\citenamefont{Kondov, McGehee,
  Zirbel, and DeMarco}}]{Kondov2011}
\bibinfo{author}{\bibfnamefont{S.~S.} \bibnamefont{Kondov}},
  \bibinfo{author}{\bibfnamefont{W.~R.} \bibnamefont{McGehee}},
  \bibinfo{author}{\bibfnamefont{J.~J.} \bibnamefont{Zirbel}},
  \bibnamefont{and} \bibinfo{author}{\bibfnamefont{B.}~\bibnamefont{DeMarco}},
  \bibinfo{journal}{Science} \textbf{\bibinfo{volume}{334}},
  \bibinfo{pages}{66} (\bibinfo{year}{2011}).

\bibitem[{\citenamefont{Jendrzejewski et~al.}(2012)\citenamefont{Jendrzejewski,
  Bernard, Mueller, Cheinet, Josse, Piraud, Pezz\'e, Sanchez-Palencia, Aspect,
  and Bouyer}}]{Jendrzejewski2012}
\bibinfo{author}{\bibfnamefont{F.}~\bibnamefont{Jendrzejewski}},
  \bibinfo{author}{\bibfnamefont{A.}~\bibnamefont{Bernard}},
  \bibinfo{author}{\bibfnamefont{K.}~\bibnamefont{Mueller}},
  \bibinfo{author}{\bibfnamefont{P.}~\bibnamefont{Cheinet}},
  \bibinfo{author}{\bibfnamefont{V.}~\bibnamefont{Josse}},
  \bibinfo{author}{\bibfnamefont{M.}~\bibnamefont{Piraud}},
  \bibinfo{author}{\bibfnamefont{L.}~\bibnamefont{Pezz\'e}},
  \bibinfo{author}{\bibfnamefont{L.}~\bibnamefont{Sanchez-Palencia}},
  \bibinfo{author}{\bibfnamefont{A.}~\bibnamefont{Aspect}}, \bibnamefont{and}
  \bibinfo{author}{\bibfnamefont{P.}~\bibnamefont{Bouyer}},
  \bibinfo{journal}{Nature Physics} \textbf{\bibinfo{volume}{8}},
  \bibinfo{pages}{398} (\bibinfo{year}{2012}).

\bibitem[{\citenamefont{Shapiro}(2012)}]{Shapiro2012}
\bibinfo{author}{\bibfnamefont{B.}~\bibnamefont{Shapiro}}, \bibinfo{journal}{J.
  Phys. A: Math. Theor.} \textbf{\bibinfo{volume}{45}}, \bibinfo{pages}{143001}
  (\bibinfo{year}{2012}).

\bibitem[{\citenamefont{Goodman}(2007)}]{Good07}
\bibinfo{author}{\bibfnamefont{J.~W.} \bibnamefont{Goodman}},
  \emph{\bibinfo{title}{Speckle phenomena in optics, Theory and applications}}
  (\bibinfo{publisher}{Roberts \& Company}, \bibinfo{year}{2007}).

\bibitem[{\citenamefont{Piraud et~al.}(2012)\citenamefont{Piraud, Aspect, and
  Sanchez-Palencia}}]{Piraud2011}
\bibinfo{author}{\bibfnamefont{M.}~\bibnamefont{Piraud}},
  \bibinfo{author}{\bibfnamefont{A.}~\bibnamefont{Aspect}}, \bibnamefont{and}
  \bibinfo{author}{\bibfnamefont{L.}~\bibnamefont{Sanchez-Palencia}},
  \bibinfo{journal}{Phys. Rev. A} \textbf{\bibinfo{volume}{85}},
  \bibinfo{pages}{063611} (\bibinfo{year}{2012}),
  \urlprefix\url{http://link.aps.org/doi/10.1103/PhysRevA.85.063611}.

\bibitem[{\citenamefont{P\l{}odzie\ifmmode~\acute{n}\else \'{n}\fi{} and
  Sacha}(2011)}]{Plodzien2011}
\bibinfo{author}{\bibfnamefont{M.}~\bibnamefont{P\l{}odzie\ifmmode~\acute{n}\else
  \'{n}\fi{}}} \bibnamefont{and}
  \bibinfo{author}{\bibfnamefont{K.}~\bibnamefont{Sacha}},
  \bibinfo{journal}{Phys. Rev. A} \textbf{\bibinfo{volume}{84}},
  \bibinfo{pages}{023624} (\bibinfo{year}{2011}),
  \urlprefix\url{http://link.aps.org/doi/10.1103/PhysRevA.84.023624}.

\bibitem[{\citenamefont{Piraud and Sanchez-Palencia}(2013)}]{Piraud2013}
\bibinfo{author}{\bibfnamefont{M.}~\bibnamefont{Piraud}} \bibnamefont{and}
  \bibinfo{author}{\bibfnamefont{L.}~\bibnamefont{Sanchez-Palencia}},
  \bibinfo{journal}{Eur. Phys. J. Special Topics}
  \textbf{\bibinfo{volume}{217}}, \bibinfo{pages}{91} (\bibinfo{year}{2013}).

\bibitem[{\citenamefont{Izrailev and Krokhin}(1999)}]{Izraivel1999}
\bibinfo{author}{\bibfnamefont{F.~M.} \bibnamefont{Izrailev}} \bibnamefont{and}
  \bibinfo{author}{\bibfnamefont{A.~A.} \bibnamefont{Krokhin}},
  \bibinfo{journal}{Phys. Rev. Lett.} \textbf{\bibinfo{volume}{82}},
  \bibinfo{pages}{4062} (\bibinfo{year}{1999}),
  \urlprefix\url{http://link.aps.org/doi/10.1103/PhysRevLett.82.4062}.

\bibitem[{\citenamefont{Kuhl et~al.}(2008)\citenamefont{Kuhl, Izrailev, and
  Krokhin}}]{Kuhl08}
\bibinfo{author}{\bibfnamefont{U.}~\bibnamefont{Kuhl}},
  \bibinfo{author}{\bibfnamefont{F.~M.} \bibnamefont{Izrailev}},
  \bibnamefont{and} \bibinfo{author}{\bibfnamefont{A.~A.}
  \bibnamefont{Krokhin}}, \bibinfo{journal}{Phys. Rev. Lett.}
  \textbf{\bibinfo{volume}{100}}, \bibinfo{pages}{126402}
  (\bibinfo{year}{2008}).

\bibitem[{\citenamefont{Paladin and Vulpiani}(1987)}]{Paladin1987}
\bibinfo{author}{\bibfnamefont{G.}~\bibnamefont{Paladin}} \bibnamefont{and}
  \bibinfo{author}{\bibfnamefont{A.}~\bibnamefont{Vulpiani}},
  \bibinfo{journal}{Phys. Rev. B} \textbf{\bibinfo{volume}{35}},
  \bibinfo{pages}{2015} (\bibinfo{year}{1987}),
  \urlprefix\url{http://link.aps.org/doi/10.1103/PhysRevB.35.2015}.

\bibitem[{\citenamefont{Piraud}(2012)}]{piraudthesis}
\bibinfo{author}{\bibfnamefont{M.}~\bibnamefont{Piraud}},
  \emph{\bibinfo{title}{Anderson localization of matter waves in correlated
  disorder : from 1D to 3D}} (\bibinfo{publisher}{PhD Thesis, Universit\'{e}
  Paris-Sud}, \bibinfo{year}{2012}).

\bibitem[{\citenamefont{Chow}(1972)}]{Chow1972}
\bibinfo{author}{\bibfnamefont{P.}~\bibnamefont{Chow}}, \bibinfo{journal}{Am.
  J. Phys.} \textbf{\bibinfo{volume}{40}}, \bibinfo{pages}{730}
  (\bibinfo{year}{1972}).

\bibitem[{\citenamefont{Lugan et~al.}(2009)\citenamefont{Lugan, Aspect,
  Sanchez-Palencia, Delande, Gr\'emaud, M\"uller, and Miniatura}}]{Luga09}
\bibinfo{author}{\bibfnamefont{P.}~\bibnamefont{Lugan}},
  \bibinfo{author}{\bibfnamefont{A.}~\bibnamefont{Aspect}},
  \bibinfo{author}{\bibfnamefont{L.}~\bibnamefont{Sanchez-Palencia}},
  \bibinfo{author}{\bibfnamefont{D.}~\bibnamefont{Delande}},
  \bibinfo{author}{\bibfnamefont{B.}~\bibnamefont{Gr\'emaud}},
  \bibinfo{author}{\bibfnamefont{C.~A.} \bibnamefont{M\"uller}},
  \bibnamefont{and}
  \bibinfo{author}{\bibfnamefont{C.}~\bibnamefont{Miniatura}},
  \bibinfo{journal}{Phys. Rev. A} \textbf{\bibinfo{volume}{80}},
  \bibinfo{pages}{023605} (\bibinfo{year}{2009}),
  \urlprefix\url{http://link.aps.org/doi/10.1103/PhysRevA.80.023605}.

\bibitem[{\citenamefont{Salasnich et~al.}(2002)\citenamefont{Salasnich, Parola,
  and Reatto}}]{salasnich02}
\bibinfo{author}{\bibfnamefont{L.}~\bibnamefont{Salasnich}},
  \bibinfo{author}{\bibfnamefont{A.}~\bibnamefont{Parola}}, \bibnamefont{and}
  \bibinfo{author}{\bibfnamefont{L.}~\bibnamefont{Reatto}},
  \bibinfo{journal}{Phys. Rev. A} \textbf{\bibinfo{volume}{65}},
  \bibinfo{pages}{043614} (\bibinfo{year}{2002}).

\bibitem[{\citenamefont{Albert et~al.}(2010)\citenamefont{Albert, Paul,
  Pavloff, and Leboeuf}}]{Albert2010}
\bibinfo{author}{\bibfnamefont{M.}~\bibnamefont{Albert}},
  \bibinfo{author}{\bibfnamefont{T.}~\bibnamefont{Paul}},
  \bibinfo{author}{\bibfnamefont{N.}~\bibnamefont{Pavloff}}, \bibnamefont{and}
  \bibinfo{author}{\bibfnamefont{P.}~\bibnamefont{Leboeuf}},
  \bibinfo{journal}{Phys. Rev. A} \textbf{\bibinfo{volume}{82}},
  \bibinfo{pages}{011602} (\bibinfo{year}{2010}),
  \urlprefix\url{http://link.aps.org/doi/10.1103/PhysRevA.82.011602}.

\bibitem[{\citenamefont{Leboeuf and Pavloff}(2001)}]{Leboeuf2001}
\bibinfo{author}{\bibfnamefont{P.}~\bibnamefont{Leboeuf}} \bibnamefont{and}
  \bibinfo{author}{\bibfnamefont{N.}~\bibnamefont{Pavloff}},
  \bibinfo{journal}{Phys. Rev. A} \textbf{\bibinfo{volume}{64}},
  \bibinfo{pages}{033602} (\bibinfo{year}{2001}),
  \urlprefix\url{http://link.aps.org/doi/10.1103/PhysRevA.64.033602}.

\bibitem[{\citenamefont{M\"uller and Delande}(2010)}]{Muller2010}
\bibinfo{author}{\bibfnamefont{C.}~\bibnamefont{M\"uller}} \bibnamefont{and}
  \bibinfo{author}{\bibfnamefont{D.}~\bibnamefont{Delande}}, in
  \emph{\bibinfo{booktitle}{Ultracold Gases and Quantum Information, Proceeding
  pf the Les Houches Summer School in Singapore 2009}}, edited by
  \bibinfo{editor}{\bibfnamefont{C.}~\bibnamefont{Miniatura}},
  \bibinfo{editor}{\bibfnamefont{L.-C.} \bibnamefont{Kwek}},
  \bibinfo{editor}{\bibfnamefont{M.}~\bibnamefont{Ducloy}},
  \bibinfo{editor}{\bibfnamefont{B.}~\bibnamefont{Gr\'emaud}},
  \bibinfo{editor}{\bibfnamefont{B.-G.} \bibnamefont{Englert}},
  \bibinfo{editor}{\bibfnamefont{A.}~\bibnamefont{Ekert}}, \bibnamefont{and}
  \bibinfo{editor}{\bibfnamefont{K.}~\bibnamefont{Phua}}
  (\bibinfo{publisher}{Oxfort University Press}, \bibinfo{year}{2010}), pp.
  \bibinfo{pages}{441--527}.

\end{thebibliography}

\end{document}